\documentclass[sigconf,nonacm,pbalance=true]{acmart}

\setcopyright{none}
\renewcommand\footnotetextcopyrightpermission[1]{}
\usepackage{xspace}

\usepackage{amsmath}
\usepackage{booktabs}
\usepackage{array}
\usepackage{tabularx}
\usepackage{enumitem}
\usepackage{graphicx}
\usepackage{xcolor}
\usepackage{microtype}
\usepackage{placeins}
\usepackage{url}
\AtBeginDocument{\setlength{\headheight}{17pt}}
\AtBeginEnvironment{thebibliography}{\scriptsize}
\AtBeginDocument{\setlength{\bibsep}{0pt plus 0.2ex}}

\newcommand{\Firm}{\textsc{Firm}}

\newcommand{\NotCompleted}{\textsc{Not-Completed}}
\newcommand{\StateResume}{\textsc{State Resume}}
\newcommand{\SegmentCommit}{\textsc{Verified Segment Commit}}
\newcommand{\MaaS}{\textsc{MaaS}}
\setlist[itemize]{leftmargin=*,nosep,topsep=2pt}
\setlist[enumerate]{leftmargin=*,nosep,topsep=2pt}
\newcolumntype{L}[1]{>{\raggedright\arraybackslash}p{#1}}
\newcolumntype{Y}{>{\raggedright\arraybackslash}X}
\makeatletter

\newcommand{\SemSpot}{\textsc{SemSpot}\xspace}
\newcommand{\mytitle}[1]{\noindent\textbf{#1}}

\title[Bridging Agent Semantics with Spot Capacity]{Bridging Agent Semantics with Spot Capacity: An Elastic and Recoverable Service Model}

\author{Minchen Yu}
\affiliation{%
  \institution{The Chinese University of Hong Kong, Shenzhen}
  \city{Shenzhen}
  \country{China}}
\email{yuminchen@cuhk.edu.cn}

\begin{document}

\begin{abstract}
LLM agents increasingly drive long-running cloud inference workloads in
which model calls differ in urgency, redundancy, completion semantics, and
replay cost. Model-as-a-Service (MaaS) platforms expose several service
models for trading cost against latency, availability, and capacity
commitment. These models operate primarily at request, job, or endpoint
scopes and provide limited support for combining transient platform supply
with the evolving semantics of an agent task.

We present \emph{\SemSpot}, a semantics-aware service model that allows
agent applications to leverage the spot capacity of LLM inference platforms.
At the request level, \SemSpot lets a provider publish short-lived offers
over successful price, completion probability, and
failure-notification deadline; the agent runtime selects among these offers
using the current task state and completion rule. An audit of 1,535 cases
from six agent benchmarks identifies four recurring workflow structures and
shows how this service model may produce different cost, service-time,
and fallback behavior.  With specialized MaaS support, Token-Level \SemSpot further preserves provider inference state and runtime-verified semantic segments inside a long request. We develop the service model, economic boundary, and the cross-layer research agenda required to realize \SemSpot.

\end{abstract}

\maketitle

\section{Introduction}

LLM agents are becoming a major source of cloud inference demand. A user objective can unfold into a long-running task that plans, retrieves information, invokes tools, explores alternatives, validates intermediate results, and revises subsequent actions. Such tasks maintain state across many model calls and reveal their execution structure online: a speculative branch may become critical after sibling failures, a verifier may initiate another round, and a join may make unfinished work irrelevant. Efficient agent serving must therefore follow the evolving structure of a live task~\cite{agentix,murakkab,parrot,sglang}.

Model-as-a-Service (MaaS) platforms provide the computational substrate for these tasks. They multiplex requests with heterogeneous context lengths and uncertain generation times over expensive accelerators while workload bursts continually change queue, compute, and memory pressure. Modern serving systems expose substantial internal flexibility through batching, disaggregation, preemption, KV-cache management, and elastic placement~\cite{sarathi,distserve,splitwise,fastserve,llumnix,conserve}. Most of this flexibility remains hidden behind a conventional inference API.

MaaS providers also offer varying service models, including Standard and Priority for interactive requests, Batch for detached work, and Flex or best-effort processing for lower-cost service~\cite{openaiFlex,googleFlex,awsTiers,openaiBatch,googleProvisioned}. These models cover useful cost--service tradeoffs, and runtimes can route different inference calls to different products. Their public contracts are generally selected for a request, job, or capacity commitment. However, they provide limited support for conditioning an offer on the current state of a live task or coordinating recovery after opportunistic execution is interrupted.

This boundary separates two views needed for efficient execution. The runtime manages the current time budget, branch redundancy, completion rule, and replay cost of each call. The platform keeps track of resource utilization, request queue, model placement, and KV-cache availability. Existing products let the runtime select a service and react to its outcome, but they do not provide a common, task-conditioned contract that relates the semantics of the current call to the available capacity. The gap becomes most visible when opportunistic work fails late, several branches fall back together, or useful in-request progress could otherwise be retained.

Cloud Spot services provide a useful economic precedent. They allow a
provider to monetize spare infrastructure while retaining a reclaim option,
and compensate the user through a discount
~\cite{awsSpot,googleSpot,spotcheck,transientGuarantees}. Flex processing has brought part of this exchange to model APIs~\cite{openaiFlex,googleFlex,awsTiers}. We argue that agent serving motivates a semantics-aware extension that connects task context with the Spot capacity of MaaS platforms. We use \emph{Firm}\footnote{\emph{Firm} denotes any protected execution path that offers sufficiently stable delivery for the selected inference unit. It may be realized by Standard or Priority request service, and serves as the stable alternative and fallback to Spot.} to denote the stable execution path against which Spot is selected and recovered.
An ideal Spot service should exchange the information needed for a joint decision. The runtime supplies only the
semantics needed to choose and recover execution---for example, the current
time budget, completion rule, fallback policy, and recovery capability. The
provider supplies a priced description of the spot capacity available for
that execution and returns a terminal outcome that the runtime can interpret
deterministically.

We call this service model \emph{\SemSpot}. \SemSpot operates at
two complementary granularities. At the \emph{request level}, the provider
publishes a short-lived menu whose entries describe a settlement horizon,
successful price, completion probability, and failure-notification deadline.
The runtime chooses among Firm execution and these Spot offers according to
the task's current state and completion rule. At the \emph{token level},
\SemSpot moves the recovery boundary inside a long request with specialized support of MaaS platforms. Provider-state
resume preserves KV-cache and decoder progress, while runtime-verified
segment commit preserves output that has already become independently valid
to the application. 

In this paper, we primarily focus on establishing and analyzing \SemSpot service model. We first model request-level \SemSpot{} under four recurring agent workflow structures identified through a preliminary audit of 1,535 benchmark cases. We then place provider operational saving and runtime bill reduction of \SemSpot to derive an economic filter and guide the \SemSpot configurations for MaaS providers. Finally, we examine the additional platform and runtime support required for token-level \SemSpot{} and clarify when physical and semantic recovery are complementary.
Realizing \SemSpot{} requires coordinated advances in three areas. A portable \emph{semantic contract} must expose enough task context and terminal information without revealing a full workflow or private infrastructure state. The \emph{agent runtime} must identify safe inference units, select among Firm and multiple Spot offers online, coordinate fan-out and cancellation, and validate recoverable progress. The \emph{MaaS platform} must calibrate offers, protect the capacity required by Firm traffic and fallback, manage KV and compute resources, and provide auditable settlement.

\noindent\textbf{Contributions.}
This paper makes the following contributions:
\begin{itemize}[leftmargin=*,itemsep=1pt,topsep=2pt]
  \item We introduce \SemSpot, a cross-layer, semantics-aware service model that connects agent task context with the spot capacity of MaaS platforms while preserving the responsibilities of the runtime and provider.
  \item We formulate request-level \SemSpot{} as a short-lived provider menu and use a preliminary audit of 1,535 cases from six agent benchmarks to derive structure-specific bill, service-time-budget, and fallback implications.
  \item We analyze operational saving and runtime bill reduction of \SemSpot for providers, yielding an economic filter and design guidelines for publishable Spot offers.
  \item We explore token-level \SemSpot{} through provider-state resume and runtime-verified segment commit, and identify the runtime and MaaS mechanisms needed to evaluate the model.
    \item We identify the open problems for semantic contracts, agent runtime, and MaaS platforms required to realize and optimize \SemSpot in practice.
\end{itemize}

\section{Background}
\subsection{Agent Workloads}
Coding, research, and data-analysis agents repeatedly alternate between model calls, tools, validators, and state updates~\cite{agentix,murakkab,parrot,sglang}. A task can contain a long causal chain, parallel candidates, an all-required join, or a feedback loop whose depth emerges only after execution. Recent workload studies also report long reusable prefixes, short outputs, and heavy-tailed generation times in coding agents~\cite{tracelab,copilotwild,agenticworkload}.
The value of a model call changes with task state. Early exploration may be redundant; the last surviving branch may become mandatory; a semantic threshold can terminate the remaining cohort; final synthesis and effect authorization require stable execution. These transitions are application semantics, not properties of a queue priority.

Long tasks can issue hundreds of inference calls, so modest per-call savings accumulate. The same elasticity can also threaten task completion. A late failure on a sequential chain consumes critical-path budget, while simultaneous misses in an all-required fan-out create a burst of Firm fallbacks. A service model for agents must therefore expose both economic opportunity and a continuation that the runtime can compose with the task.

\begin{table*}[t]
  \caption{Comparison of customer-visible MaaS service models.
  \emph{Operational unit} denotes the granularity of execution and recovery
  semantics, rather than the billing meter. For \SemSpot{}, ``token'' refers
  to request-internal inference progress.}
  \label{tab:services}
  \centering
  \small
  \begin{tabularx}{\textwidth}{
      @{}
      L{2.25cm}
      L{2.15cm}
      L{4.45cm}
      Y
      @{}}
    \toprule
    Service model
      & Operational unit
      & Cost--service exchange
      & Runtime-visible outcome / recovery \\
    \midrule

    Standard / Priority
      & Request
      & Stable or premium online execution
      & Complete response or ordinary service error \\

    Batch
      & Detached job
      & Lower price for a longer asynchronous completion window
      & Per-member completion status; unfinished members may be retried \\

    Flex / best effort
      & Request
      & Lower price for weaker latency or capacity availability
      & Complete response or generic failure; timeout, retry, and fallback
        remain client-managed \\

    Provisioned
      & Capacity envelope
      & Capacity commitment for predictable throughput or dedicated service
      & Ordinary request outcomes under a reservation, quota, or overflow
        policy \\

    \midrule

    \textbf{\SemSpot{} (proposed)}
      & \textbf{Request / token}
      & Task-conditioned, discounted, and revocable execution, with optional
        request-internal recovery
      & Request level: complete response or actionable
        \NotCompleted{} with a defined Firm fallback; token level:
        compatible state resume and/or a runtime-verified segment \\

    \bottomrule
  \end{tabularx}
\end{table*}

\subsection{MaaS Service Models}

A customer-visible MaaS service model determines three aspects of
execution. The \emph{operational unit} specifies the granularity over which
service and recovery semantics apply, such as a request, a detached job, a
capacity envelope, or request-internal token progress. The
\emph{cost--service exchange} specifies what service property is traded for
a lower price or stronger capacity guarantee. The
\emph{runtime-visible outcome} determines what the runtime can observe and
how it can continue after completion, rejection, or interruption.
Operational unit is distinct from the billing meter: input, cached-input,
and output tokens may remain the metered quantities for every service model.

Table~\ref{tab:services} summarizes the main MaaS service models
~\cite{openaiFlex,googleFlex,awsTiers,openaiBatch,googleProvisioned}.
Standard and Priority provide stable request execution; Batch discounts
detached jobs with a longer completion window; Flex and best-effort services
discount request-level execution while relaxing latency or capacity
availability; and provisioned products reserve an aggregate capacity
envelope.

SemSpot builds on the same basic economic exchange as Flex---a lower price
in return for weaker execution assurance---but makes two additional
capabilities explicit. At the \emph{request level}, the provider publishes
a short-lived, task-composable offer and settles the attempt as either a
complete response or an actionable \NotCompleted{} outcome, enabling a
defined Firm fallback. At the \emph{token level}, SemSpot moves the recovery
boundary inside the request and may expose compatible provider state, a
runtime-verified output segment, or both. The distinction from Flex is
therefore not the availability of discounted best-effort inference itself;
it is the combination of a task-conditioned offer, an explicit
continuation, and optional finer-grained recovery.

\section{\SemSpot}
\label{sec:model}
\subsection{Definition and Properties}

\mytitle{Definition.}
\SemSpot{} is a customer-visible MaaS service model that exposes transient inference capacity through discounted, revocable offers with explicit continuations. The runtime associates compact task constraints and a fallback or recovery policy with an inference unit. The platform prices, admits, and executes that unit on Spot capacity. Every attempt resolves to an outcome from which the runtime can proceed deterministically.

\mytitle{Cross-layer service abstraction.}
The public contract joins two complementary control planes. The runtime contributes constraints that affect service selection, including urgency, completion rule, fallback budget, and requested recovery capability. The platform contributes a priced execution offer, a terminal outcome, and any retained inference state. Prompts, full task graphs, queue state, topology, and raw KV layouts remain private to their owners.

\mytitle{Semantics-aware orchestration.}
With \SemSpot, Agent runtime can compose multiple service models according to semantics at a fine granularity. It chooses a service for each inference unit and revises later choices after verifier results, sibling failures, joins, or budget updates. This granularity lets the runtime protect mandatory work, exploit application-owned redundancy, and cancel work whose marginal value has vanished.

\mytitle{Recoverable interruption.}
A rejected or reclaimed attempt returns a defined continuation. Request-level \SemSpot{} yields a complete response or an actionable failure followed by one Firm retry. Token-level \SemSpot{} can additionally return compatible model state or identify a runtime-verified output boundary. The runtime validates task correctness; the platform controls physical execution, reclaim, and state retention.

The abstraction creates a common optimization surface. An agent runtime can reduce the bill of elastic work while preserving Firm execution for critical stages, and can route online among multiple Spot offers as task state changes. A provider can monetize transient queue, compute, and memory headroom, reclaim it to protect Firm demand, and use compact continuation information to limit destructive preemption and uncontrolled fallback bursts. Standardized terminal and recovery semantics also make heterogeneous implementations comparable.

\subsection{Scope and Concepts}

We clarify the scope and concepts for later discussion.

\mytitle{Task.}
An \emph{agent task} is a runtime-owned workflow containing inference calls, tools, validators, and state transitions. The runtime owns the task graph, completion conditions, and all authority over external effects.

\mytitle{Inference unit.}
An \emph{inference unit} $u$ is one immutable model call directly served by MaaS. It is identified by (i) the input snapshot consumed by the call, including the prompt, context, and task-state version, and (ii) the model and inference configuration, such as model version, reasoning effort, and decoding settings. Changing either part creates a new unit. A task can contain many units separated by runtime-owned tools or validation.

\mytitle{Attempt.}
An \emph{attempt} is one submission of $u$ under one service mode. The same unit may first run on \SemSpot and later on Firm. A separate execution epoch is unnecessary unless a future service changes price, protection, or authority inside one live attempt.

\mytitle{Recovery boundary.}
A \emph{recovery boundary} identifies the latest state from which continuation is correct and economically useful. Request-level Spot returns to the original immutable input. Token-level Spot can add provider KV cache state, a runtime-verified output segment, or both.

\subsection{Two Execution Granularities}
\mytitle{Request-level \SemSpot.}
The request-level profile treats one inference unit as atomic. A successful attempt returns a complete response at the quoted price. An unsuccessful attempt returns an actionable failure, after which the runtime immediately retries the same unit on Firm. This profile has a compact public contract and applies when whole-request replay is safe and affordable.

\mytitle{Token-level \SemSpot.}
The token-level profile moves the recovery boundary inside a long request and therefore requires specialized MaaS support. \StateResume{} retains or exports compatible KV-cache, decoder position, and serving metadata. \SegmentCommit{} preserves a closed output segment after a runtime-owned verifier establishes that it is independently reusable. The capabilities address different forms of replay and can be combined.

Both profiles require an unambiguous terminal outcome and a continuation understood by the runtime. \SemSpot{} does not make the provider responsible for task correctness, nor does it recover arbitrary tools or external effects. 
Instead, the provider expose sufficient information for the runtime to evaluate efficient and recoverable execution.

\section{Request-Level \SemSpot}
\label{sec:request}
\subsection{Contract and Single-Unit Analysis}
\label{sec:single}
Request-level \SemSpot treats one inference unit as atomic. The first profile rests on four assumptions.

\mytitle{(A1) Equivalent and replay-safe unit.}
The spot attempt and its Firm fallback execute the same immutable input and task-state version under the same model, reasoning profile, decoding and safety contract, and complete-response boundary. Replaying the unit does not repeat an external effect.

\mytitle{(A2) Atomic visibility.}
The runtime consumes only a complete response. Partial output cannot trigger tools, mutate shared state, or authorize an external effect before the response is accepted.

\mytitle{(A3) Zero-charge unsuccessful attempt.}
An attempt that ends as \NotCompleted{} is not billed. This assumption keeps the first-order model transparent; a product that charges incomplete work requires an additional settlement term.

\mytitle{(A4) Immediate one-shot Firm fallback.}
A \NotCompleted{} outcome immediately sends the same unit to Firm, and the runtime does not issue another spot attempt. Capacity shortage is a common source of rejection or reclaim, so repeated spot attempts can encounter correlated scarcity while consuming the time reserved for completion. One-shot fallback bounds retry amplification and guarantees a stable continuation.

\mytitle{Provider menu.}
For unit $u$, the provider publishes a short-lived menu
\begin{equation}
  \mathcal M(u)=\{o_j=(h_j,r_j,p_j,f_j)\}_{j=1}^{m}.
  \label{eq:menu}
\end{equation}
The \emph{horizon} $h_j$ is the latest time at which the Spot attempt settles. The \emph{price ratio} $r_j$ is the successful Spot price divided by the Firm price. The \emph{completion probability} $p_j$ is the probability of a complete response by $h_j$ for the provider's current request class. The \emph{failure-notification deadline} $f_j$ is a product SLO, for example, ``99\% of unsuccessful attempts are reported within $f_j$.'' A provider without an early-failure commitment sets $f_j=h_j$.

\mytitle{Service-time budget.}
The timing fields allocate inference service time to one unit. They do not predict its realized duration or the task's end-to-end latency. Before issuing $u$, the runtime reserves time for tools, human interaction, downstream inference, validation, and finalization, and assigns the remaining amount as the unit's \emph{Spot slack} $S_u$. Let $L_F(u)$ be the Firm service-time budget for the same unit. A menu point is feasible under the conservative check
\begin{equation}
  \max\{h_j,\;f_j+L_F(u)\}\le S_u.
  \label{eq:feasible}
\end{equation}
The successful path fits within $h_j$; the failure path receives a notification by $f_j$ and then one Firm budget. The provider absorbs request-length, batching, and queue uncertainty into the calibrated menu, while the runtime recomputes $S_u$ after every accepted result, join, or feedback iteration.

\mytitle{Bill notation.}
For any fixed execution scope $A$---one inference unit, one workflow stage, or one paired evaluation window---let $B_F(A)$ denote the customer bill when exactly the same work uses Firm, and let $B_S(A)$ denote the bill under a candidate Spot policy. This common notation keeps the single-unit, workflow-level, and provider-accounting analyses on the same monetary baseline.

\mytitle{Single-unit bill.}
Under (A3)--(A4), a successful Spot attempt costs a fraction $r$ of the Firm charge, while an unsuccessful attempt is free and the Firm fallback pays the full Firm charge. Hence
\begin{equation}
  \frac{\mathbb E[B_S(u)]}{B_F(u)}
  =pr+(1-p)=1-p(1-r).
  \label{eq:usercost}
\end{equation}
Thus $p(1-r)$ is the expected bill reduction as a fraction of the Firm bill for the same unit. A smaller failure deadline $f$ does not change Equation~\ref{eq:usercost}; it preserves more task budget for the Firm path. A longer horizon can raise $p$, yet it also consumes more slack and can delay failure. The menu exposes this tradeoff rather than fixing one global best-effort tier.

\subsection{Workflow-Level Analysis}
\label{sec:workflow}

\begin{figure*}[t]
  \centering
  \includegraphics[width=0.94\textwidth]{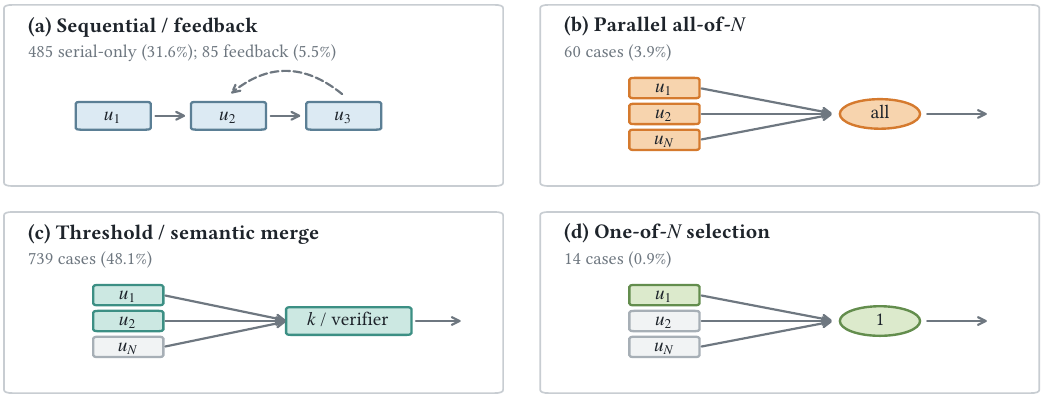}
  \Description{A two-by-two figure shows the topology and preliminary benchmark coverage of sequential or feedback execution, all-of-N joins, threshold or semantic merge, and one-of-N selection.}
  \caption{Four task completion structures found in the benchmark audit. Counts and percentages are independent preliminary labels over 1,535 cases and can overlap.}
  \label{fig:workflows}
\end{figure*}

Equation~\ref{eq:usercost} describes one inference unit. The task completion rule determines how many Spot misses require Firm repair and how service-time uncertainty composes across units. We audited all primary cases in BrowseComp-Plus, Long-Horizon Terminal-Bench, LongDS-Bench, $\tau^2$-bench, MCPMark, and Terminal-Bench 2.0, for 1,535 cases total~\cite{browsecompplus,lhtb,longds,taubench,mcpmark,terminalbench}. The outcome-blind static characterization remains preliminary and subject to independent adjudication. Figure~\ref{fig:workflows} shows four recurring structures and their independent coverage.

\mytitle{Analysis assumptions.}
The main comparison keeps the application fixed: \textsc{All-Firm} and \SemSpot{} use the same task graph, inference units, launch order, verifier, and cancellation rule. To isolate the completion rule, the closed-form bill expressions make two simplifying assumptions: branches are homogeneous and share the same $(p,r)$, and parallel Spot outcomes are independent, so the number of completed branches is $X\sim\mathrm{Binomial}(N,p)$. Real branches can differ in length and experience correlated scarcity; their bill and service-time tails must be measured from joint completion traces. The timing discussion below concerns the inference service-time budget allocated to the stage, not the task's end-to-end latency. For any scalar $z$, $(z)^+=\max\{z,0\}$, so $(k-X)^+$ is the number of results still missing after $X$ completions. The formulas assume simultaneous launch and omit early cancellation.

\mytitle{(1) Sequential and feedback chains.}
The corpus contains 485 strict serial-only cases (31.6\%) and 85 feedback cases (5.5\%, an overlapping label). If $m$ ordered units use the same Spot point, the normalized expected bill is
\begin{equation}
  R^{B}_{\mathrm{seq}}
  =\frac{\mathbb E[B_S(\mathcal G_{\mathrm{seq}})]}{B_F(\mathcal G_{\mathrm{seq}})}
  =rp+(1-p)=1-p(1-r).
  \label{eq:seqratio}
\end{equation}
Failures normally arrive one at a time, so fallback concurrency is small. Their service times accumulate along the critical path. With heterogeneous units, the conservative budget is the sum of the selected per-unit budgets $\max\{h_i,f_i+L_{F,i}\}$. Feedback depth is unknown online, so the runtime replans after each iteration and moves later units to Firm as the remaining budget contracts.

\mytitle{(2) All-of-$N$ joins.}
The audit identifies 60 all-required cases (3.9\%). Every miss must be repaired, and the normalized expected bill again reduces to
\begin{equation}
  R^{B}_{\mathrm{all}}
  =\frac{\mathbb E[B_S(\mathcal G_{\mathrm{all}})]}{B_F(\mathcal G_{\mathrm{all}})}
  =rp+\frac{\mathbb E[N-X]}{N}
  =rp+(1-p).
  \label{eq:allratio}
\end{equation}
The cost ratio matches the sequential case, while the time behavior differs. The stage completes at the slowest usable branch, and Spot misses can release several Firm fallbacks at the same join. Heterogeneous branch lengths and shared fallback queues create a long-tail maximum that must be measured jointly. Under the illustrative homogeneous independent model and no time for fallback, stage completion is $p^N$; a 99\% all-of-eight target requires per-unit $p\ge99.874\%$. All-required stages therefore need high completion probability, cohort limits, and explicit Firm reserve.

\mytitle{(3) Threshold and semantic merge.}
The audit identifies 739 semantic or threshold merges (48.1\%). Suppose the same $N$ branches run under both policies and the verifier accepts after $k$ useful results. The normalized expected Spot bill is
\begin{equation}
  R^{B}_{k,N}
  =\frac{\mathbb E[B_S(\mathcal G_{k,N})]}{B_F(\mathcal G_{k,N})}
  =rp+\frac{\mathbb E[(k-X)^+]}{N}.
  \label{eq:kratio}
\end{equation}
The second term is the expected Firm \emph{top-up} needed after the Spot horizon, normalized by the original $N$-call All-Firm bill. It is no larger than $1-p$ because $(k-X)^+\le N-X$; the completion rule absorbs part of the Spot miss whenever $k<N$. The service-time behavior follows the $k$th usable-result order statistic rather than the slowest branch. A semantic verifier can therefore cut off the long tail once sufficiency, diversity, or another completion predicate holds, and should cancel remaining Spot and Firm work immediately. Increasing $N$ beyond the original task graph is a separate hedging policy and requires a separate baseline.

\mytitle{(4) One-of-$N$ selection.}
The audit identifies 14 one/select stages (0.9\%). The same $N$ candidates are launched in both policies. Ignoring early cancellation, the normalized expected Spot bill is
\begin{equation}
  R^{B}_{1,N}
  =\frac{\mathbb E[B_S(\mathcal G_{1,N})]}{B_F(\mathcal G_{1,N})}
  =rp+\frac{(1-p)^N}{N}.
  \label{eq:oneratio}
\end{equation}
The last term has a direct meaning: all $N$ Spot candidates miss with probability $(1-p)^N$, which triggers one Firm fallback; division by $N$ normalizes that repair against the same $N$-candidate All-Firm workflow. The stage service time is governed by the first \emph{usable} candidate, so redundancy can suppress the tail even when individual branches are heterogeneous. Service completion alone is insufficient---the native verifier determines whether a candidate is usable. Early cancel-on-accept can further reduce both bills, and its exact value must be measured with paired completion and validation traces.

The workflow analysis therefore adds information that the single-unit model cannot capture. Sequential stages accumulate failure delay; all-required stages expose the maximum and synchronized fallback burst; threshold stages replace the maximum with a $k$th order statistic and reduce Firm top-up; selection stages replace it with the first accepted result. These distinctions determine which menu points remain useful after task-level time and capacity constraints are applied.

\subsection{Designing A Viable \SemSpot{} Menu}
The single-unit and workflow analyses characterize the runtime side of the service. The platform must also determine whether a candidate menu point creates enough operational saving while preserving \Firm{} objectives. We first measure the saving created by the platform policy, and then choose a price that divides it between the runtime and the platform.

\mytitle{Operational saving.}
Fix a paired evaluation window $W$ in which the all-\Firm{} reference and the candidate \SemSpot{} policy see the same arrival trace, task graph, model mix, capacity trace, and \Firm{} SLO. Let $B_F(W)$ be the all-\Firm{} customer bill, and let $K_F(W)$ and $K_S(W)$ be the platform's aggregate operating cost under the two executions. To keep the notation compact, we write them as $B_F$, $K_F$, and $K_S$ below. The spot-policy cost includes incomplete work, \Firm{} fallback, reserved headroom, state handling, and measured interference with protected traffic. Batching, locality, and queue interactions are captured by paired replay or controlled deployment rather than assigned to an artificial per-request GPU cost.

We define the normalized operational saving as
\begin{equation}
  v=\frac{K_F-K_S}{B_F}.
  \label{eq:value}
\end{equation}
Thus, $v$ measures how much platform cost the candidate policy removes, expressed as a fraction of the same all-\Firm{} bill used by the runtime-side analysis. A policy with $v\le0$ creates no efficiency that can be shared through price.

\noindent\textbf{Price allocation.}
For a homogeneous class with completion probability $p$, successful price ratio $r$, and zero-charge unsuccessful attempts, the runtime receives an expected bill reduction of $p(1-r)$ on the same normalized baseline. The platform retains the remainder, $v-p(1-r)$, as incremental profit. Both sides gain when
\begin{equation}
  0<p(1-r)<v,
  \label{eq:winwin}
\end{equation}
which yields the price interval
\begin{equation}
  \max\!\left\{0,1-\frac{v}{p}\right\}<r<1.
  \label{eq:priceinterval}
\end{equation}
The lower boundary is the deepest successful-attempt discount that the measured operational saving can support. For heterogeneous workloads, the runtime saving becomes a bill-weighted average, while the same accounting identity applies.

\begin{figure}[t]
  \centering
  \includegraphics[width=\columnwidth]{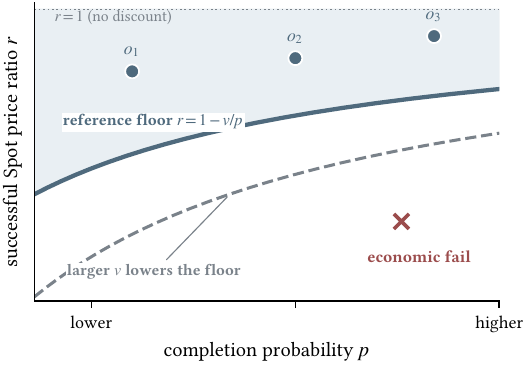}
  \Description{A schematic p-r plane shows economic price floors for a SemSpot menu. A larger normalized operational saving v shifts the floor downward. Candidate menu points above the applicable floor are economically feasible, while a point below it is labeled economic fail.}
  \caption{Economic filtering of a request-level \SemSpot{} menu. Each horizon and platform policy produces a candidate $(p,r)$ point and a measured normalized operational saving $v$. A point is economically publishable only above $r=1-v/p$; task-time, workflow, and \Firm{}-protection constraints are applied separately. Curves and points are schematic.}
  \label{fig:economics}
\end{figure}

Figure~\ref{fig:economics} maps the accounting test back to the provider menu. The solid curve is the price floor for a reference policy, and the dashed curve shows how a larger $v$ permits a deeper discount. A point below its applicable floor gives the runtime more discount than the platform policy creates in operational saving. Points above the floor remain candidates rather than automatic choices: the runtime must still evaluate the local time budget and completion rule, and the platform must still preserve fallback capacity and \Firm{} non-interference.

\noindent\textbf{Takeaways.} We summarize three takeaways. (1) The platform can increase $v$ through system-level optimization, including efficient scheduling, resource scaling, fast rejection, efficient KV management, and tighter fallback reservation, to reduce $K_S$ relative to $K_F$. A larger $v$ directly widens the space for a useful discount.
(2) Completion probability $p$ and price ratio $r$ must be \emph{designed jointly}. Equation~\ref{eq:priceinterval} gives their qualitative coupling for a measured $v$: a menu point with a different horizon or control policy changes both $p$ and the price floor.
(3) Economic feasibility is \emph{necessary but not sufficient}. A low $p$, late failure notification, correlated fallback, or an insufficient task-time budget can make a point unusable even when it lies above the economic floor. The final decision must also satisfy Equation~\ref{eq:feasible}, the workflow-specific tail behavior, fallback-capacity limits, and \Firm{}-service protection.

\section{Token-Level \SemSpot}
\label{sec:token}
\subsection{Motivation and scope}
Request-level \SemSpot{} treats an inference unit as atomic. After interruption, the runtime retries the unit from its original input on \Firm{}. This boundary is simple and broadly applicable, but it can repeat a long prefill and substantial decode work. TraceLab reports coding-agent generation latency around 5.7~s at P50, 22.2~s at P90, and 1.4~min at P99, with median prefixes around 115K--126K tokens and a reported P90 above 467K tokens for one agent family~\cite{tracelab}. A production GitHub Copilot study reports median prompt, cached-prompt, and output sizes of roughly 68K, 63K, and 247 tokens~\cite{copilotwild}.

Token-level \SemSpot{} moves the recovery boundary inside a request. Input, cached-input, and output tokens remain the billing meter; ``token-level'' refers to the granularity of preserved execution progress. This profile requires specialized platform support for retaining and restoring model state, as well as runtime support for binding recovered progress to the exact input and inference configuration. We consider two complementary capabilities: \emph{state resume} preserves physical inference progress, while \emph{verified segment commit} preserves application-valid semantic progress.

\subsection{State Resume}
State resume extends familiar KV and prefix reuse across a spot-to-\Firm{} transition. When a spot attempt is interrupted, the platform may retain or export prefix KV, generated-token KV, decoder position, and the serving metadata needed to continue the same inference unit. The public capability should expose only what the runtime needs: a state handle, its compatibility scope, lifetime, and restore outcome. Internal KV layout, paging, compression, transfer, and victim selection remain platform choices.

A usable handle must bind the exact input snapshot and inference configuration, including model and version, tokenizer, quantization, adapter, region, and security domain. The runtime rejects a handle that no longer matches the selected \Firm{} continuation. The service must also define deterministic behavior for early eviction, failed restore, destination unavailability, and unused state after normal completion.

State retention is not free capacity. Retained KV and decoder state compete with active prefill, decode, memory, and transfer bandwidth. The platform must decide which states to keep, where to place them, and when to evict or export them while preserving \Firm{} headroom. Existing paging, migration, and preemptive-serving mechanisms can implement these actions; \SemSpot{} makes the continuation and its observable semantics customer-visible.

\subsection{Verified Segment Commit}
State resume answers where model execution can continue, but an opaque state handle does not identify which partial outputs are already valid to the application. Agent workflows may need progress that survives the loss of provider-local state, can be consumed by a downstream stage, or can move to another execution context. Verified segment commit addresses this semantic boundary by recording a closed output object whose validity is established by the runtime.

This complements state resume in three conditions. (1) Physical continuation does not certify application results. A KV cache or decoder handle cannot establish that partial output is complete, correct, or safe for downstream use. It cannot by itself represent a verified evidence item, an accepted patch, or a completed branch.
(2) Physical state and semantic results have different lifetimes. A handle may expire or become incompatible after a model, region, quantization, adapter, or provider change. A closed, version-bound semantic object can remain useful after the physical state disappears.
(3) The two capabilities remove different replay. State resume preserves physical inference work; verified segment commit preserves application-valid work. Their combination, aligned recovery, binds the latest verified segment to compatible KV or decoder state and avoids both model recomputation and regeneration or revalidation of accepted results.

\mytitle{Examples.}
We illustrate the applicability of verified segment commit with two examples. 
Following prior benchmarks~\cite{browsecompplus,xgrammar,jsonschemabench}, the first example is a fixed-corpus request that emits one JSONL evidence record per line, with fields for a claim, source document, cited span, and relation. After each record closes, the runtime checks the schema, corpus revision, and cited span. If interruption occurs after four verified records, those records can feed downstream synthesis or a new request even if the state handle is unavailable. State resume can additionally avoid replaying the long prefix when a compatible handle survives.

A second example is a long request that emits several complete patch candidates. If candidate~1 passes native tests before later generation is interrupted, state resume can continue the original request when the handle remains valid, whereas verified segment commit lets the runtime preserve and use the accepted patch even after the handle expires or execution moves elsewhere.

Verified segment commit should be omitted when useful internal boundaries are rare, validation is expensive or non-monotone, or request splitting with prompt caching reaches the same frontier. Both capabilities require explicit compatibility, lifetime, terminal-result, and settlement semantics.

\section{Open Problems}
\label{sec:open}

\SemSpot{} connects two systems with complementary knowledge: the agent
runtime understands task semantics, whereas the \MaaS{} platform controls
resource supply and physical execution. The central research question is
how the two sides can exchange the minimum information required for
service selection and recovery, without exposing a complete task graph or
private platform state.

\subsection{Agent Runtime}
\mytitle{Task decomposition and spot eligibility.}
The runtime must decompose a dynamic task into immutable inference units and
identify which units can safely use a spot offer. This decision requires more
than detecting model calls: the runtime must determine the input and
task-state version consumed by each call, the completion rule of the
surrounding stage, and whether partial output can affect shared state or
external systems. Request-level \SemSpot{} applies only when a failed spot
attempt can be retried from the same input without duplicating an effect;
token-level \SemSpot{} additionally requires a valid internal recovery
boundary. A central research problem is how to infer these properties from
agent programs, schemas, validators, and tool interfaces.

\mytitle{Semantics-aware online routing.}
For every eligible inference unit, the runtime must choose among \Firm{},
multiple spot offers, deferred execution, and token-level recovery. The
choice depends on the current service-time budget, the unit's position in the
task, its completion rule, replay cost, available fallback capacity, and the
uncertainty of the provider's offer. These quantities change after verifier
outcomes, sibling failures, joins, and feedback iterations, making routing a
constrained online-control problem rather than a one-time workload
classification. An important research question is which task semantics
provide measurable benefit beyond client-side history and simple timeout or
retry policies.

\mytitle{Workflow-level group control.}
Inference units within the same task cannot always be routed independently.
For example, an all-of-$N$ workflow stage should choose cohort width and reserve
sufficient \Firm{} fallback capacity for simultaneous misses.
The research opportunity is to lift serving control from per-request
priority to task-level completion semantics. A compact representation of
cohort identity, completion rule, fallback limit, and cancellation condition
could enable structure-aware routing without exposing the complete workflow.
The runtime must also prevent low-priced spot execution from creating
unnecessary fan-out, redundant paid work, or synchronized \Firm{} fallback.

\mytitle{Recovery validation.}
After a spot interruption, the runtime must decide whether to retry the
entire inference unit on \Firm{}, resume provider state, reuse a verified
segment, or combine the latter two. A resume handle is valid only when its
input, model, tokenizer, adapter, quantization, region, and security scope
remain compatible with the continuation. A semantic segment is reusable
only when it is closed, version-bound, independently validated, and not
invalidated by later task updates.
This creates a research problem at the boundary between systems recovery and
application correctness. The runtime needs a unified policy that compares
whole-request replay, physical state restoration, and semantic reuse while
accounting for validation cost and stale-state risk. Multi-provider
execution makes this harder: verified semantic objects may be portable,
whereas raw KV state will often remain provider-specific.

\subsection{MaaS Platform}

\mytitle{Offer construction.}
The platform should construct short-lived \SemSpot menu based on time-varying resource availability.
This is an online systems and product-design problem rather than static
pricing. The platform must decide how long an offer remains valid, how to
express uncertainty or confidence, and when a class is too unstable to
publish. It should expose only non-dominated menu points whose completion
behavior is calibrated and whose measured saving remains positive after
failed work, fallback, reserved headroom, and \Firm{} interference are
included.

\mytitle{Joint admission and fallback protection.}
Spot admission cannot be controlled independently of \Firm{} service. A rise
in protected demand may reclaim many spot attempts, which can then return
immediately as \Firm{} fallback traffic. This feedback loop is especially
strong for all-required cohorts, while threshold and selection stages may
need only a limited top-up or no fallback after their completion condition is
met.
The platform therefore needs admission policies that jointly reserve
capacity for protected arrivals, active spot work, cleanup, and potential
fallback. 

\mytitle{Cluster scheduling and resource management.}
\SemSpot{} provides the platform with task-level semantics that are absent
from ordinary request priorities, such as the remaining service-time budget,
completion rule, branch redundancy, fallback requirement, and recovery
cost. The platform can combine these semantics with its private view of
resource availability to coordinate admission, placement, batching,
priority, and reclaim across both cluster-management and serving-control
loops. The goal is to complete more task-relevant work with transient
capacity while preserving \Firm{} service.
This cross-layer design raises several research questions. The platform must
translate compact and potentially uncertain task semantics into robust
scheduling decisions. The runtime supplies
constraints and decision-relevant hints, while the platform retains final
authority over physical scheduling, resource allocation, and reclaim.

\mytitle{KV cache and state management.}
Token-level \SemSpot{} turns KV cache and runtime state from an internal
optimization into a customer-visible recovery capability. The platform must
define stable state identity, compatibility scope, retention time, and
restore outcome, while remaining free to choose the underlying paging,
compression, placement, or migration mechanism. A state object may be kept
in GPU memory, offloaded to host memory, transferred to another worker, or
evicted with explicit absence.
Retained state competes with active inference for memory and bandwidth, so
preserving every interrupted request can reduce the very capacity that spot
execution aims to expose. The key research question is when state reuse creates enough
task-level value to justify its resource and management cost.

\mytitle{Observability and evaluation.}
Evaluating the service model requires paired runtime and platform traces
under the same arrivals, resource pressure, and task events. Metrics should
include task success, end-to-end completion, local service-time feasibility,
effective bill, fallback count and burst, \Firm{} interference, and recovery-state overhead. Such
counterfactual evaluation is necessary to separate the value of the
\SemSpot{} contract from the quality of a particular scheduler or
workload-specific implementation.

\section{Related Work and Limitations}
\mytitle{LLM serving and state management.}
JITServe and uncertainty-aware output-length work refine decisions with incomplete request information~\cite{jitserve,uncertainlength}. Sarathi-Serve, DistServe, Splitwise, Torpor, FaaScale, and JANUS improve goodput, elasticity, or resource decomposition~\cite{sarathi,distserve,splitwise,torpor,faascale,janus}. SpotServe, FastServe, ConServe, Llumnix, and vLLM provide preemption, paging, migration, and state-management substrates~\cite{spotserve,fastserve,conserve,llumnix,vllm}. These mechanisms can implement spot execution and state recovery; \SemSpot{} specifies how such capabilities are exposed and composed as a service.

\mytitle{Agent execution and semantic recovery.}
Parrot, SGLang, Agentix, and Murakkab use program structure or semantic variables to optimize agent execution~\cite{parrot,sglang,agentix,murakkab}. ExoFlow and DART study workflow durability and semantic recovery~\cite{exoflow,dart}. Model routers and cascades choose capability and quality~\cite{frugalgpt,routellm}; they complement \SemSpot{} routing over execution service and recovery. \SemSpot{} exports a compact subset of task constraints through a priced interface while leaving task correctness and effect authority in the runtime.

\mytitle{Limitations.}
\SemSpot{} is useful only when normalized operational saving $v$ is positive and sufficiently stable. High \Firm{} utilization can leave little spot capacity; incomplete work, fallback reserve, cleanup, and state retention can consume the saving. Token-level recovery can similarly collapse to ordinary caching or request splitting when state is large, incompatible, or cheaper to recompute. The benchmark audit is preliminary, and the economic analysis remains a counterfactual design framework until validated with production traces and controlled MaaS experiments.

\section{Conclusion}
Agent workflows expose task semantics that are not directly reflected in
existing MaaS service contracts. We proposed \SemSpot{}, a semantics-aware
service model that connects these semantics with transient platform capacity
through request-level spot offers and optional in-request recovery. Our
analysis shows that its value depends on how workflow structure shapes
fallback behavior and whether the platform creates sufficient operational
saving to sustain the offered discount. This framing turns spot execution
from a generic low-price option into a cross-layer service decision, and
provides a basis for future runtime and platform designs.

\bibliographystyle{ACM-Reference-Format}
\bibliography{references}

@misc{openaiFlex,
  author = {{OpenAI}},
  title = {Flex Processing},
  year = {2026},
  howpublished = {OpenAI API Documentation},
  url = {https://developers.openai.com/api/docs/guides/flex-processing},
  note = {Accessed 2026-08-20}
}

@misc{openaiBatch,
  author = {{OpenAI}},
  title = {Batch API},
  year = {2026},
  howpublished = {OpenAI API Documentation},
  url = {https://platform.openai.com/docs/guides/batch},
  note = {Accessed 2026-08-20}
}

@misc{googleFlex,
  author = {{Google}},
  title = {Flex Inference},
  year = {2026},
  howpublished = {Gemini API Documentation},
  url = {https://ai.google.dev/gemini-api/docs/flex-inference},
  note = {Preview; accessed 2026-08-20}
}

@misc{awsTiers,
  author = {{Amazon Web Services}},
  title = {Service Tiers for Optimizing Performance and Cost},
  year = {2026},
  howpublished = {Amazon Bedrock Documentation},
  url = {https://docs.aws.amazon.com/bedrock/latest/userguide/service-tiers-inference.html},
  note = {Accessed 2026-08-20}
}

@misc{googleProvisioned,
  author = {{Google Cloud}},
  title = {Provisioned Throughput for Generative AI},
  year = {2026},
  howpublished = {Vertex AI Documentation},
  url = {https://cloud.google.com/vertex-ai/generative-ai/docs/provisioned-throughput/overview},
  note = {Accessed 2026-08-20}
}

@misc{awsSpot,
  author = {{Amazon Web Services}},
  title = {Amazon EC2 Spot Instances},
  year = {2026},
  howpublished = {AWS Documentation},
  url = {https://docs.aws.amazon.com/AWSEC2/latest/UserGuide/using-spot-instances.html},
  note = {Accessed 2026-08-20}
}

@misc{googleSpot,
  author = {{Google Cloud}},
  title = {Spot VMs},
  year = {2026},
  howpublished = {Google Cloud Documentation},
  url = {https://cloud.google.com/compute/docs/instances/spot},
  note = {Accessed 2026-08-20}
}

@inproceedings{spotcheck,
  author = {Prateek Sharma and David Irwin and Prashant Shenoy},
  title = {SpotCheck: Designing a Derivative IaaS Cloud on the Spot Market},
  booktitle = {Proceedings of the Tenth European Conference on Computer Systems (EuroSys)},
  year = {2015}
}

@inproceedings{transientGuarantees,
  author = {Supreeth Subramanya and Amr Rizk and David Irwin},
  title = {Cloud Spot Markets Are Not Sustainable: The Case for Transient Guarantees},
  booktitle = {8th USENIX Workshop on Hot Topics in Cloud Computing (HotCloud)},
  year = {2016},
}

@article{browsecompplus,
  author = {Zijian Chen and Xueguang Ma and Shengyao Zhuang and Ping Nie and Kai Zou and Andrew Liu and Joshua Green and Kshama Patel and Ruoxi Meng and Mingyi Su and Sahel Sharifymoghaddam and Yanxi Li and Haoran Hong and Xinyu Shi and Xuye Liu and Nandan Thakur and Crystina Zhang and Luyu Gao and Wenhu Chen and Jimmy Lin},
  title = {BrowseComp-Plus: A More Fair and Transparent Evaluation Benchmark of Deep-Research Agent},
  journal = {arXiv preprint arXiv:2508.06600},
  year = {2025}
}

@article{lhtb,
  author = {Zongxia Li and Zhongzhi Li and Yucheng Shi and Ruhan Wang and Junyao Yang and Zhichao Liu and Xiyang Wu and Anhao Li and Yue Yu and Ninghao Liu and Lichao Sun and Haotao Mi and Leowei Liang},
  title = {Long-Horizon-Terminal-Bench: Testing the Limits of Agents on Long-Horizon Terminal Tasks with Dense Reward-Based Grading},
  journal = {arXiv preprint arXiv:2607.08964},
  year = {2026}
}

@article{longds,
  author = {Kewei Xu and Xiaoben Lu and Shuofei Qiao and Zihan Ding and Haoming Xu and Lei Liang and Ningyu Zhang},
  title = {LongDS-Bench: On the Failure of Long-Horizon Agentic Data Analysis},
  journal = {arXiv preprint arXiv:2605.30434},
  year = {2026}
}

@inproceedings{taubench,
  author = {Shunyu Yao and Noah Shinn and Pedram Razavi and Karthik Narasimhan},
  title = {$\tau$-Bench: A Benchmark for Tool-Agent-User Interaction in Real-World Domains},
  booktitle = {International Conference on Learning Representations (ICLR)},
  year = {2025}
}

@article{mcpmark,
  author = {Zijian Wu and Xiangyan Liu and Xinyuan Zhang and Lingjun Chen and Fanqing Meng and Lingxiao Du and Yiran Zhao and Fanshi Zhang and Yaoqi Ye and Jiawei Wang and Zirui Wang and Jinjie Ni and Yufan Yang and Arvin Xu and Michael Qizhe Shieh},
  title = {MCPMark: A Benchmark for Stress-Testing Realistic and Comprehensive MCP Use},
  journal = {arXiv preprint arXiv:2509.24002},
  year = {2025}
}

@article{terminalbench,
  author = {Mike A. Merrill and Alexander G. Shaw and Nicholas Carlini and others},
  title = {Terminal-Bench: Benchmarking Agents on Hard, Realistic Tasks in Command Line Interfaces},
  journal = {arXiv preprint arXiv:2601.11868},
  year = {2026}
}

@article{dart,
  author = {Ke Yang and Panpan Li and Zonghan Wu and Kejin Xu and Huaxi Huang and Xiaoshui Huang},
  title = {DART: Semantic Recoverability for Structured Tool Agents},
  journal = {arXiv preprint arXiv:2605.23311},
  year = {2026}
}

@inproceedings{vllm,
  author = {Woosuk Kwon and Zhuohan Li and Siyuan Zhuang and Ying Sheng and Lianmin Zheng and Cody Hao Yu and Joseph E. Gonzalez and Hao Zhang and Ion Stoica},
  title = {Efficient Memory Management for Large Language Model Serving with PagedAttention},
  booktitle = {Proceedings of the 29th ACM Symposium on Operating Systems Principles (SOSP)},
  year = {2023},
  pages = {611--626}
}

@inproceedings{sarathi,
  author = {Amey Agrawal and Nitin Kedia and Ashish Panwar and Jayashree Mohan and Nipun Kwatra and Bhargav Gulavani and Alexey Tumanov and Ramachandran Ramjee},
  title = {Taming Throughput-Latency Tradeoff in LLM Inference with Sarathi-Serve},
  booktitle = {18th USENIX Symposium on Operating Systems Design and Implementation (OSDI)},
  year = {2024},
  pages = {117--134}
}

@inproceedings{distserve,
  author = {Yinmin Zhong and Shengyu Liu and Junda Chen and Jianbo Hu and Yibo Zhu and Xuanzhe Liu and Xin Jin and Hao Zhang},
  title = {DistServe: Disaggregating Prefill and Decoding for Goodput-Optimized Large Language Model Serving},
  booktitle = {18th USENIX Symposium on Operating Systems Design and Implementation (OSDI)},
  year = {2024},
  pages = {193--210}
}

@inproceedings{splitwise,
  author = {Pratyush Patel and Esha Choukse and Chaojie Zhang and Aashaka Shah and Inigo Goiri and Saeed Maleki and Ricardo Bianchini},
  title = {Splitwise: Efficient Generative LLM Inference Using Phase Splitting},
  booktitle = {51st Annual International Symposium on Computer Architecture (ISCA)},
  year = {2024}
}

@article{llumnix,
  author = {Biao Sun and Ziming Huang and Hanyu Zhao and Wencong Xiao and Xinyi Zhang and Yong Li and Wei Lin},
  title = {Llumnix: Dynamic Scheduling for Large Language Model Serving},
  journal = {arXiv preprint arXiv:2406.03243},
  year = {2024}
}

@inproceedings{spotserve,
  author = {Xupeng Miao and Chunan Shi and Jiangfei Duan and Xiaoli Xi and Dahua Lin and Bin Cui and Zhihao Jia},
  title = {SpotServe: Serving Generative Large Language Models on Preemptible Instances},
  booktitle = {Proceedings of the 29th ACM International Conference on Architectural Support for Programming Languages and Operating Systems (ASPLOS)},
  year = {2024},
  pages = {1112--1127}
}

@inproceedings{fastserve,
  author = {Bingyang Wu and Yinmin Zhong and Zili Zhang and Shengyu Liu and Fangyue Liu and Yuanhang Sun and Gang Huang and Xuanzhe Liu and Xin Jin},
  title = {FastServe: Iteration-Level Preemptive Scheduling for Large Language Model Inference},
  booktitle = {23rd USENIX Symposium on Networked Systems Design and Implementation (NSDI)},
  year = {2026},
  pages = {57--74}
}

@article{conserve,
  author = {Yifan Qiao and Shu Anzai and Shan Yu and Haoran Ma and Shuo Yang and Yang Wang and Miryung Kim and Yongji Wu and Yang Zhou and Jiarong Xing and Joseph E. Gonzalez and Ion Stoica and Harry Xu},
  title = {ConServe: Fine-Grained GPU Harvesting for LLM Online and Offline Co-Serving},
  journal = {arXiv preprint arXiv:2410.01228},
  year = {2025}
}

@inproceedings{torpor,
  author = {Minchen Yu and Ao Wang and Dong Chen and Haoxuan Yu and Xiaonan Luo and Zhuohao Li and Wei Wang and Ruichuan Chen and Dapeng Nie and Haoran Yang and Yu Ding},
  title = {Torpor: GPU-Enabled Serverless Computing for Low-Latency, Resource-Efficient Inference},
  booktitle = {2025 USENIX Annual Technical Conference (USENIX ATC)},
  year = {2025},
  pages = {597--613}
}

@inproceedings{faascale,
  author = {Minchen Yu and Rui Yang and Chaobo Jia and Zhaoyuan Su and Sheng Yao and Tingfeng Lan and Yuchen Yang and Zirui Wang and Yue Cheng and Wei Wang and Ao Wang and Ruichuan Chen},
  title = {FaaScale: Unlocking Fast LLM Scaling for Serverless Inference},
  booktitle = {Proceedings of Machine Learning and Systems (MLSys)},
  year = {2026}
}

@article{janus,
  author = {Zhexiang Zhang and Ye Wang and Yumiao Zhao and Jiayu Xiao and Qianjing Yang and Xiangyu Wang and Jingzhe Jiang and Qizhen Weng and Ruichuan Chen and Shaohuai Shi and Adel N. Toosi and Yin Chen and Minchen Yu},
  title = {JANUS: Disaggregating Attention and Experts for Scalable MoE Inference},
  journal = {arXiv preprint arXiv:2512.13525},
  year = {2026}
}

@inproceedings{parrot,
  author = {Chaofan Lin and Zhenhua Han and Chengruidong Zhang and Yuqing Yang and Fan Yang and Chen Chen and Lili Qiu},
  title = {Parrot: Efficient Serving of LLM-Based Applications with Semantic Variable},
  booktitle = {18th USENIX Symposium on Operating Systems Design and Implementation (OSDI)},
  year = {2024},
  pages = {929--945}
}

@inproceedings{sglang,
  author = {Lianmin Zheng and Liangsheng Yin and Zhiqiang Xie and Chuyue Sun and Jeff Huang and Cody Hao Yu and Shiyi Cao and Christos Kozyrakis and Ion Stoica and Joseph E. Gonzalez and Clark Barrett and Ying Sheng},
  title = {SGLang: Efficient Execution of Structured Language Model Programs},
  booktitle = {Advances in Neural Information Processing Systems},
  volume = {37},
  year = {2024},
  pages = {62557--62583}
}

@inproceedings{agentix,
  author = {Michael Luo and Xiaoxiang Shi and Colin Cai and Tianjun Zhang and Justin Wong and Yichuan Wang and Chi Wang and Yanping Huang and Zhifeng Chen and Joseph E. Gonzalez and Ion Stoica},
  title = {Agentix: An Efficient Serving Engine for LLM Agents as General Programs},
  booktitle = {23rd USENIX Symposium on Networked Systems Design and Implementation (NSDI)},
  year = {2026},
  pages = {2443--2459}
}

@inproceedings{murakkab,
  author = {Gohar Irfan Chaudhry and Esha Choukse and Haoran Qiu and Inigo Goiri and Rodrigo Fonseca and Adam Belay and Ricardo Bianchini},
  title = {Murakkab: Resource-Efficient Agentic Workflow Orchestration in Cloud Platforms},
  booktitle = {20th USENIX Symposium on Operating Systems Design and Implementation (OSDI)},
  year = {2026},
  pages = {567--587}
}

@inproceedings{exoflow,
  author = {Siyuan Zhuang and Stephanie Wang and Eric Liang and Yi Cheng and Ion Stoica},
  title = {ExoFlow: A Universal Workflow System for Exactly-Once DAGs},
  booktitle = {17th USENIX Symposium on Operating Systems Design and Implementation (OSDI 23)},
  pages = {269--286},
  year = {2023},
  publisher = {USENIX Association},
}

@article{tracelab,
  author = {Kan Zhu and Mathew Jacob and Chenxi Ma and Yi Pan and Stephanie Wang and Arvind Krishnamurthy and Baris Kasikci},
  title = {TraceLab: Characterizing Coding Agent Workloads for LLM Serving},
  journal = {arXiv preprint arXiv:2606.30560},
  year = {2026},
}

@article{copilotwild,
  author = {Banruo Liu and Haoran Qiu and {\'I}{\~n}igo Goiri and Rodrigo Fonseca and Ricardo Bianchini and Esha Choukse},
  title = {Agentic Coding in the Wild: Characterizing GitHub Copilot Traces at Production Scale},
  journal = {arXiv preprint arXiv:2608.00101},
  year = {2026},
}

@article{agenticworkload,
  author = {Yichao Yuan and Ankita Nayak and Souvik Kundu and Nishil Talati},
  title = {Agentic AI Workload Characteristics},
  journal = {arXiv preprint arXiv:2605.26297},
  year = {2026},
}

@inproceedings{jitserve,
  author = {Wei Zhang and Zhiyu Wu and Yi Mu and Rui Ning and Banruo Liu and Nikhil Sarda and Myungjin Lee and Fan Lai},
  title = {JITServe: SLO-aware LLM Serving with Imprecise Request Information},
  booktitle = {23rd USENIX Symposium on Networked Systems Design and Implementation (NSDI 26)},
  year = {2026},
  pages = {825--848},
  publisher = {USENIX Association},
}

@article{uncertainlength,
  author = {Haoyu Zheng and Yongqiang Zhang and Fangcheng Fu and Xiaokai Zhou and Hao Luo and Hongchao Zhu and Yuanyuan Zhu and Hao Wang and Xiao Yan and Jiawei Jiang},
  title = {Scheduling LLM Inference with Uncertainty-Aware Output Length Predictions},
  journal = {arXiv preprint arXiv:2604.00499},
  year = {2026},
}

@article{xgrammar,
  author = {Yixin Dong and Charlie F. Ruan and Yaxing Cai and Ruihang Lai and Ziyi Xu and Yilong Zhao and Tianqi Chen},
  title = {XGrammar: Flexible and Efficient Structured Generation Engine for Large Language Models},
  journal = {arXiv preprint arXiv:2411.15100},
  year = {2025},
}

@article{jsonschemabench,
  author = {Saibo Geng and Hudson Cooper and Micha{\l} Moskal and Samuel Jenkins and Julian Berman and Nathan Ranchin and Robert West and Eric Horvitz and Harsha Nori},
  title = {Generating Structured Outputs from Language Models: Benchmark and Studies},
  journal = {arXiv preprint arXiv:2501.10868},
  year = {2025},
}

@article{routellm,
  author = {Isaac Ong and Amjad Almahairi and Vincent Wu and Wei-Lin Chiang and Tianhao Wu and Joseph E. Gonzalez and M. Waleed Kadous and Ion Stoica},
  title = {RouteLLM: Learning to Route LLMs with Preference Data},
  journal = {arXiv preprint arXiv:2406.18665},
  year = {2024},
}

@article{frugalgpt,
  author = {Lingjiao Chen and Matei Zaharia and James Zou},
  title = {FrugalGPT: How to Use Large Language Models While Reducing Cost and Improving Performance},
  journal = {arXiv preprint arXiv:2305.05176},
  year = {2023},
}
\end{document}